\documentclass[conference]{IEEEtran}

\usepackage{cite}
\usepackage{graphicx}
\usepackage{epstopdf}
\usepackage{amsfonts,amsmath,amssymb}
\usepackage{graphicx}
\usepackage{bm,bbm}
\usepackage{subcaption}
\usepackage{stfloats}
\usepackage{algorithm}
\usepackage{algpseudocode}
\newfloat{algorithm}{t}{lop}

\let\ss= \scriptscriptstyle

\newcommand{\s}{\mathrm{s}}
\newcommand{\m}{\mathrm{m}}
\newcommand{\radX}[1]{r_\mathrm{#1}}
\newcommand{\radR}{\radX{r}}
\newcommand{\radD}{\radX{0}}
\newcommand{\rx}{\mathrm{r}}
\newcommand{\hit}{\mathrm{hit}}
\newcommand{\apriori}{\emph{a priori}}

\newcommand{\RMC}{\textnormal{RMC}}
\newcommand{\APMC}{\textnormal{APMC}}

\begin{document}

\title{A New Simulation Algorithm for Absorbing Receiver in Molecular Communication}

\author{\IEEEauthorblockN{Yiran Wang$^\dag$, Adam Noel$^\ddag$, and Nan Yang$^\dag$}
\IEEEauthorblockA{$^\dag$Research School of Engineering, Australian National University, Canberra, ACT 2601, Australia\\
$^\ddag$School of Engineering, University of Warwick, Coventry, CV4 7AL, UK}
Email: yiran.wang@anu.edu.au, adam.noel@warwick.ac.uk, nan.yang@anu.edu.au}

\markboth{Submitted to IEEE SECON 2018 Workshop}{Wang \MakeLowercase{\textit{et
al.}}: A New Simulation Algorithm for Absorbing Receivers in Molecular Communication}

\maketitle

\begin{abstract}
The simulation of diffusion-based molecular communication systems with absorbing receivers often requires a high computational complexity to produce accurate results. In this work, a new  \textit{a priori} Monte Carlo (APMC) algorithm is proposed to precisely simulate the molecules absorbed at a spherical receiver when the simulation time step length is relatively large. This algorithm addresses the limitations of the current refined Monte Carlo (RMC) algorithm, since the RMC algorithm provides accurate simulation only for a relatively small time step length. The APMC algorithm is demonstrated to achieve a higher simulation efficiency than the existing algorithms by finding that the APMC algorithm, for a relatively large time step length, absorbs the fraction of molecules expected by analysis, while other algorithms do not.
\end{abstract}

\section{Introduction}\label{sec:intro}

Molecular communication (MC) has recently emerged as an underpinning paradigm of exchanging information in nano-scale environments such as organs, tissues, soil, and water~\cite{nakanobook}. This information exchange is conducted among nanomachines which are the devices with nano-scale functional components. Within the research community, diffusion-based MC is a simple but commonly adopted MC system. In such a system, information molecules propagate using kinetic energy only, which preserves a high energy efficiency. One of the major challenges in designing and analyzing a diffusion-based MC system is receiver modeling. The majority of the existing MC studies have considered two types of receivers: passive receivers and active receivers. Passive receivers do not impose any impact on molecule propagation, while active receivers can absorb molecules when they hit the receiver's surface. In nature, most receiver nanomachines react with or remove selected information molecules from the environment when they hit the receiver's surface. Thus, active receivers are generally more realistic than passive ones in describing the chemical detection mechanism in MC. This motivates us to investigate the properties of absorbing receivers.

The notion of diffusion with absorption is a long existing phenomenon that has been described in the literature, e.g., \cite{berg1993random}. Considering a diffusion-based MC system with a single fully absorbing receiver within an unbounded three-dimensional environment, \cite{Yilmaz_CL_2014} presented the hitting rate of molecules at different times and the fraction of molecules absorbed until a given time. 
Recently, \cite{Deng_TMBMC_2015} and \cite{ahmadzadeh2016comprehensive} evaluated the impact of receiver with reversible absorption on the performance of MC systems, where a molecule can be released back to the environment at some time after being captured by the receiver.

Apart from theoretical analysis such as  \cite{Yilmaz_CL_2014,Deng_TMBMC_2015,ahmadzadeh2016comprehensive}, the simulation of MC systems is also an effective means for performance evaluation. One of the most common MC simulation methods is particle-based microscopic simulation. For diffusion-based MC, molecules are moved by adding Gaussian random variables (RVs) to their $x$-, $y$-, and $z$-coordinates at the end of every simulation time step. In practice, molecules may actually diffuse into an absorbing receiver between two sampling times. To determine this absorption, some existing simulation algorithms simply compared the observed coordinates of molecules with those of the receiver~\cite{yilmaz2014simulation,Deng_TMBMC_2015,llatser2014n3sim}. As a consequence, the molecules that hit a receiver between sampling times cannot be considered as ``absorbed'' by using these simulation algorithms. Recently, \cite{noel2017simulating} and \cite{arifler2017} investigated the possibility of molecule absorption between sampling times. Specifically, \cite{noel2017simulating} declared a molecule as ``absorbed'' if its straight-line trajectory within a time step crossed an absorbing surface. Differently, \cite{arifler2017} approximated the intra-step absorption probability for spherical receiver boundaries using the equation for planar receiver boundaries given by \cite{andrews2004stochastic}. This approximation was referred to as the refined Monte Carlo (RMC) algorithm.

The key contribution in this paper is that we propose a new algorithm, i.e., the \textit{a priori} Monte Carlo (APMC) algorithm, for simulating an MC system with an absorbing receiver, considering a relatively large time step length. We show that using our APMC algorithm, the fraction of molecules absorbed at the receiver precisely matches the corresponding analytical result when the time step length is relatively large, or equivalently, the receiver's radius is small. This demonstrates that our APMC algorithm achieves a high simulation efficiency.


\section{System Model}\label{sec:system_model}

\begin{figure}[!t]
  \centering
  \includegraphics[width=0.4\textwidth]{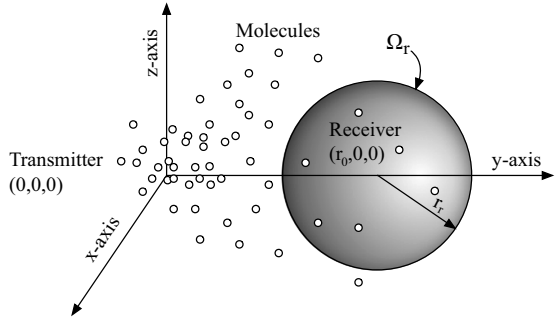}
\caption{Illustration of our system model. The TX is a point transmitter located at $(0,0,0)$, the RX is a spherical receiver located at $(\radD,0,0)$ with $\radR$ being the radius and $\Omega_\textnormal{r}$ being the RX's fully absorbing boundary. Molecules propagate in the environment according to Brownian motion.\vspace{-1em}}\label{MC system illustration}
\end{figure}

\begin{figure}[!t]
  \centering
  \includegraphics[width=0.31\textwidth]{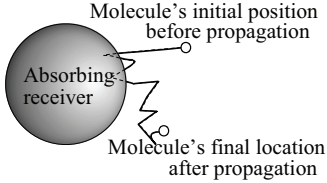}
\caption{Illustration of the intra-step molecule movement. There is a possibility that a molecule crossed an absorbing boundary within one time step, even if its initial and final positions during that time step are both outside the absorbing receiver.\vspace{-1em}}\label{intra-step absorption}
\end{figure}

We consider a diffusion-based MC system within a three-dimensional space, as depicted in Fig.~\ref{MC system illustration}. In this system, a point transmitter (TX) is located at the origin of the space and a single fully absorbing spherical receiver (RX) is centered at location $(\radD,0,0)$. We denote $\radR$ as the RX's radius and $\Omega_{\rx}$ as the RX's boundary. At the beginning of a transmission process, the TX instantaneously releases $N$ molecules. We assume that the molecules are small enough to be considered as points. Once released, molecules diffuse in the environment according to Brownian motion until hitting the RX's boundary. We denote $N_{\hit}\left(\Omega_{\rx},t|\radD\right)$ as the number of molecules released from the origin at time $t_\textnormal{0} = 0\,\s$ and absorbed by the RX at time $t$. As per \cite[Eq.~(3.116)]{schulten2000lectures}, we express $N_{\hit}(\Omega_{\rx},t|\radD)$ as
\begin{equation}\label{yilmazequation}
N_{\hit}\left(\Omega_{\rx},t|\radD\right)=
\frac{N\radR}{\radD}\textnormal{erfc}\left(\frac{\radD-\radR}{\sqrt{4Dt}}\right),
\end{equation}
where $D$ is the diffusion coefficient and $\textnormal{erfc}\left(\cdot\right)$ is the complementary error function. Here, $D$ describes the proportionality constant between the flux due to molecular diffusion and the gradient in the concentration of molecules.

As mentioned in Section~\ref{sec:intro}, the majority of the existing simulation algorithms for absorbing RXs did not consider the possibility of \emph{intra-step molecule absorption}. This absorption is depicted in Fig.~\ref{intra-step absorption}, which shows that the actual trajectory of a molecule may cross the absorbing boundary during one simulation time step, even if its initial position at the beginning of the time step and its final position at the end of the same time step are both outside the absorbing RX. If this crossing occurs, the molecule is absorbed by the RX in practice. The ignorance of this absorption leads to an underestimation of the number of molecules absorbed, thus deteriorating the accuracy of simulation. In this work, we refer to the probability that a molecule is absorbed during a simulation time step as the \emph{intra-step absorption probability}. We note that the intra-step absorbing probability was only considered in~\cite{arifler2017,noel2017simulating}. In \cite{arifler2017}, the intra-step absorption probability of a fully absorbing RX with the spherical boundary was approximated as that of a fully absorbing RX with an infinite \emph{planar} boundary, given by~\cite[Eq. (10)]{andrews2004stochastic}
\begin{align}\label{Andrews2004}
{\textrm{Pr}}_{\ss\RMC}=\exp{\left(-\frac{l_\textrm{i}l_\textrm{f}}{D\Delta t}\right)},
\end{align}
where $l_\textrm{i}$ is the initial distance of a molecule from the absorbing boundary at the beginning of a time step, $l_\textrm{f}$ is the final distance of a molecule from the absorbing boundary at the end of the same time step, and $\Delta t$ is the simulation time step length.

\section{Simulation Algorithms}\label{sim}

In this section, we first clearly present the common structure of the existing simulation algorithms for absorbing receivers. Then we present our proposed APMC algorithm.

\subsection{Common Structure of Existing Simulation Algorithms}

Built upon the observation of the existing simulation algorithms for microscopic molecule absorption (such as those in~\cite{yilmaz2014simulation,noel2017simulating,arifler2017}), we find that they follow a common structure. This common structure is presented in \textbf{Algorithm~\ref{globalalgorithm}}.


\begin{algorithm}[!t]
\caption{The common structure of simulation algorithms for a single absorbing RX} \label{globalalgorithm}
\begin{algorithmic}[1]
\State Determine the end time of simulation.
\ForAll {simulation time steps}
\If {$t=0$}
\State Add $N$ molecules to environment.
\EndIf
\State Scan all \emph{not-yet-absorbed} molecules, i.e., molecules which are not absorbed by the RX.
\ForAll {\emph{not-yet-absorbed} molecules}
\State Propagate each molecule for one step according to Brownian motion.\label{alg:normalRV1}
\State Determine if the molecule is absorbed. \label{alg:absorption}molecules absorbed
\State Update the location and status of the molecule.
\EndFor
\State Record the number of molecules absorbed by the RX.
\EndFor
\end{algorithmic}
\end{algorithm}

We clarify that each algorithm has its own criterion for determining whether or not a molecule is absorbed by the RX, as given in Line~\ref{alg:absorption} of \textbf{Algorithm~\ref{globalalgorithm}}. The determination criteria for the existing algorithms in~\cite{yilmaz2014simulation,noel2017simulating,arifler2017} are summarized as follows:
\begin{itemize}
\item
As per the determination criterion in \cite{yilmaz2014simulation}, the molecules being observed inside the RX at the end of a time step are absorbed. This criterion is referred to as the simplistic Monte Carlo (SMC) algorithm.
\item
As per the determination criterion in \cite{noel2017simulating}, a molecule is absorbed if the line segment from its initial position to its final position crosses the RX's boundary. However, we note that the line segment crossing the RX's surface is neither sufficient nor necessary to correctly detect intra-step absorption.
\item
As per the determination criterion in \cite{arifler2017}, referred to as the RMC algorithm, \eqref{Andrews2004} is used to calculate the intra-step absorption probability of a fully absorbing RX with the spherical boundary.
\end{itemize}

\subsection{A New A Priori Monte Carlo Algorithm}

In this subsection, we propose a new simulation algorithm for approximating the fraction of molecules absorbed at a fully absorbing RX when $\sqrt{D\Delta t}/\radR$ is larger than that considered for the RMC algorithm in \cite{arifler2017}. We refer to the newly proposed algorithm as the APMC algorithm.

The procedure of the APMC algorithm is to first calculate, \emph{before} the $j$th molecule diffuses, the probability that this molecule will be absorbed in the \emph{current} time step. This probability depends on the distance between this molecule and the center of the RX, $d_{j}$, and the time step length, $\Delta{t}$. Specifically, this probability is calculated as
\begin{equation}\label{analysisequationtimestep}
{\textrm{Pr}}_{\ss\APMC}=\frac{\radR}{d_j}\textrm{erfc}\left(\frac{d_j-\radR}{\sqrt{4D\Delta t}}\right),
\end{equation}
which is obtained by scaling \eqref{yilmazequation} by $N$, replacing the total simulation time $t$ with $\Delta t$, and replacing $\radD$ with $d_j$. Then the molecule absorption is determined by generating a uniform RV $u$, where $0\leq{}u\leq1$, and comparing its value with the probability obtained by \eqref{analysisequationtimestep}. A molecule is marked as ``absorbed'' if $u\leq{\textrm{Pr}}_{\ss\APMC}$. After determining the molecules absorbed, each not-yet-absorbed molecule is propagated according to Brownian motion. If any not-yet-absorbed molecule is inside the RX's boundary at the end of the current time step, we \emph{revert} the movement of this molecule and let it propagate again, until this molecule diffuses to a location outside the RX. The APMC algorithm is detailed in \textbf{Algorithm \ref{algorithmapriori}}.

\begin{algorithm}[!t]
\caption{The APMC algorithm for molecule absorption}\label{algorithmapriori}
\begin{algorithmic}[1]
\State Determine the end time of simulation.
\ForAll {simulation time steps}
\If {$t = 0$}
\State Release $N$ molecules into environment.
\EndIf
\State Scan all \emph{not-yet-absorbed} molecules.
\ForAll {\emph{not-yet-absorbed} molecules}
\State Calculate the distance between the $j$th molecule to $(\radD,0,0)$, denote by $d_j$.
\State Calculate the absorbed probability $\textnormal{Pr}_{\ss\APMC}$ for each \emph{not-yet-absorbed} molecule using \eqref{analysisequationtimestep} with $\radR$, $d_j$, $D$, and $\Delta t$.
\If { $\textnormal{Pr}_{\ss\APMC}\geq u$} \label{alg:aprioriabsorptionline}
    \State The molecule is absorbed.
    \EndIf
    \EndFor
\ForAll {\emph{not-yet-absorbed} molecules}
\State Propagate the molecule for one step. \label{alg:APMCRV1}
\EndFor
\ForAll {free molecules in environment}
\While {the molecule's distance to $(\radD,0,0)\leq\radR$}
    \State Revert the movement of this molecule.
    \State Propagate this molecule again. \label{alg:APMCRV2}
\EndWhile
\EndFor
\State Record the number of molecules absorbed at the RX after this time step.
\EndFor
\end{algorithmic}
\end{algorithm}

\section{Numerical Results}\label{sec:numerical}

In this section, we use Monte Carlo simulations to compare the time-varying results produced by the SMC algorithm, the RMC algorithm, and the APMC algorithm for a single receiver. The algorithms are implemented in MATLAB. Throughout this section, we set the diffusion coefficient as $D=10^{-9}\,{\m^2}/{\s}$ and assume that other parameters vary. In the figures, $M$ denotes the number of time steps. If not otherwise noted, the TX-RX distance is $\radD = 50\,\mu\m$ and the number of molecules released is $N = 10^6$. The fraction of molecules absorbed is defined as the ratio between the number of molecules absorbed at the RX and the total number of molecules.

\begin{figure}[!t]
\centering
    \begin{subfigure}[b]{0.24\textwidth}
    \includegraphics[width=\textwidth,height=2in]{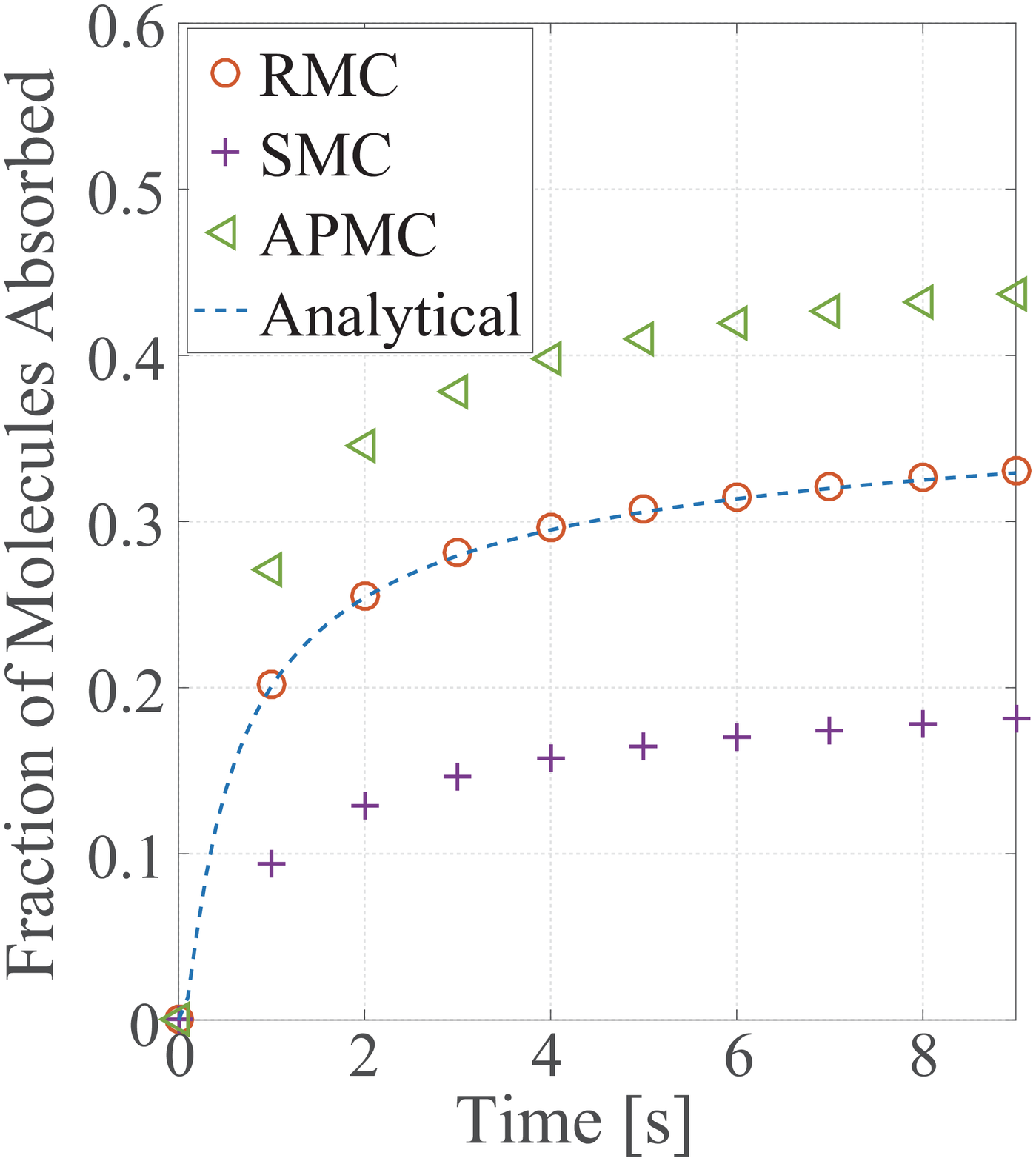}
    \caption{$\radR = 20\,\mu \m$}\end{subfigure}
    \begin{subfigure}[b]{0.24\textwidth}
    \includegraphics[width=\textwidth,height=2in]{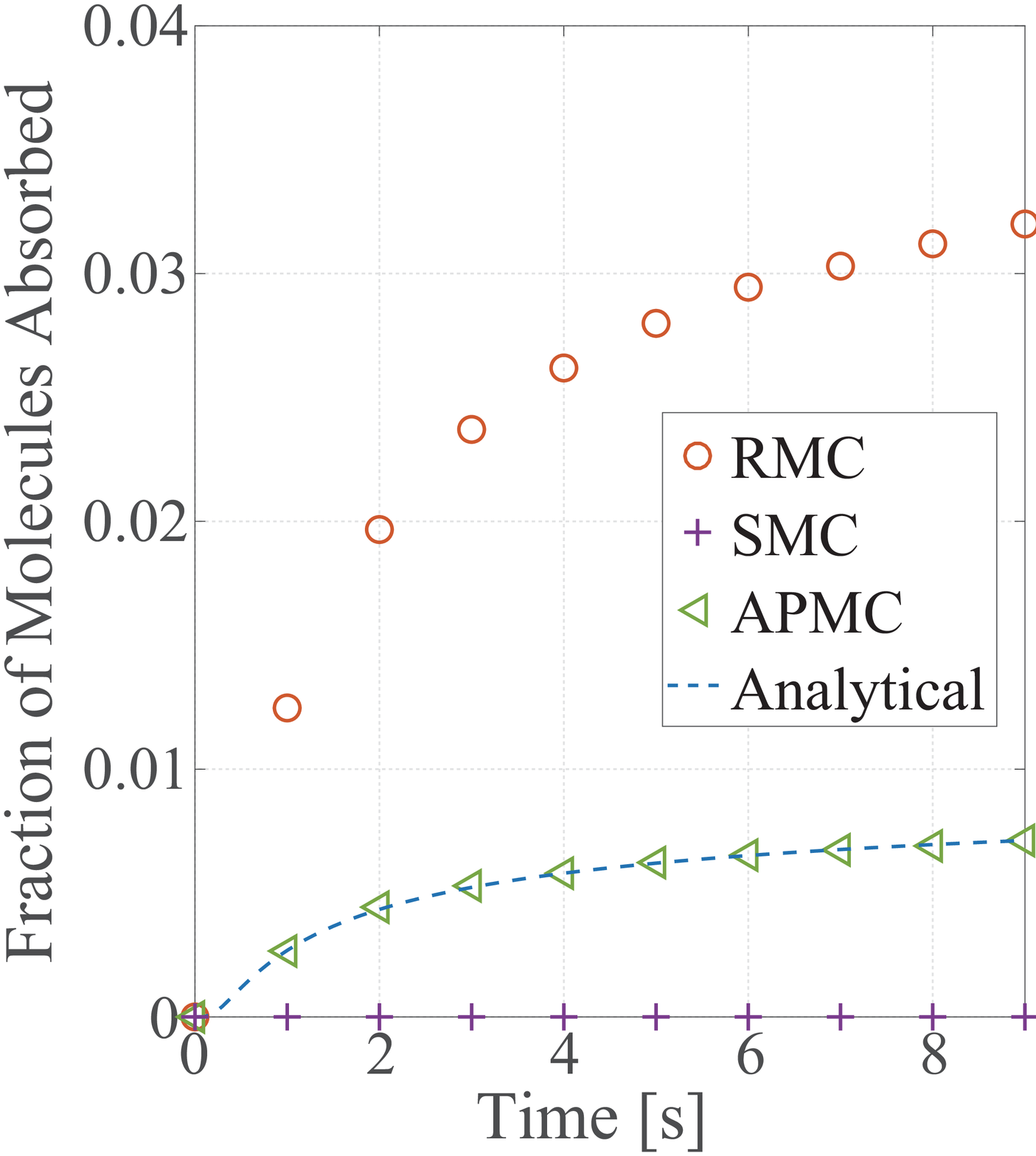}
    \caption{$\radR = 0.5\,\mu \m$}\end{subfigure}
    \caption{Comparison of the time-varying results produced by the SMC algorithm, the RMC algorithm, and the APMC algorithm for $M = 100$ and $\Delta t = 0.1\,\s$.\vspace{-1em}}\label{timevarying1}
\end{figure}

Fig.~\ref{timevarying1} plots the simulated fraction of molecules absorbed versus time $t$ for $M = 100$, $\Delta t = 0.1\,\s$, and $\radR = 20\,\mu \m$ or $0.5\,\mu \m$. In this figure, the analytical result obtained from \eqref{yilmazequation} is also plotted for evaluating the accuracy of the algorithms. From this figure, we observe that our APMC algorithm achieves a higher accuracy when $\radR$ decreases while $\Delta t$, $D$, and $\radD$ remain unchanged. Specifically, Fig.~\ref{timevarying1}(a) shows that the RMC algorithm matches the fraction of molecules absorbed expected by the analytical result when $\sqrt{D \Delta t}/\radR$ is small, which meets our expectation. Indeed, the performance of the RMC algorithm depends on the value of $\sqrt{D \Delta t}/\radR$. When $\sqrt{D \Delta t}/\radR$ approaches 0 and $\radD$ is larger than $\radR$, the surface area of the RX can be approximated by an infinite plane. Therefore, the probability of a molecule entering the RX between sampling times is comparable to the probability of a molecule crossing a planar boundary between sampling times. We also observe from Fig.~\ref{timevarying1}(a) that when $\sqrt{D \Delta t}/\radR$ is small, the SMC algorithm and the APMC algorithm underestimates and overestimates the fraction of molecules absorbed, respectively. Unlike Fig.~\ref{timevarying1}(a), Fig.~\ref{timevarying1}(b) shows that the APMC algorithm matches the fraction of molecules absorbed expected by the analytical result whereas the other two algorithms do not. Particularly, the fractions of molecules absorbed produced by the RMC and SMC algorithms are very far away from the analytical result when $t>0\,\s$. Indeed, when $\sqrt{D \Delta t}/\radR$ is large, the spherical RX's boundary cannot be approximated by a plane. Therefore, the RMC algorithm overestimates the fraction of molecules absorbed.

\begin{figure}[!t]
\centering
    \begin{subfigure}[b]{0.24\textwidth}\label{planar_0106}
    \includegraphics[width=\textwidth,height=2in]{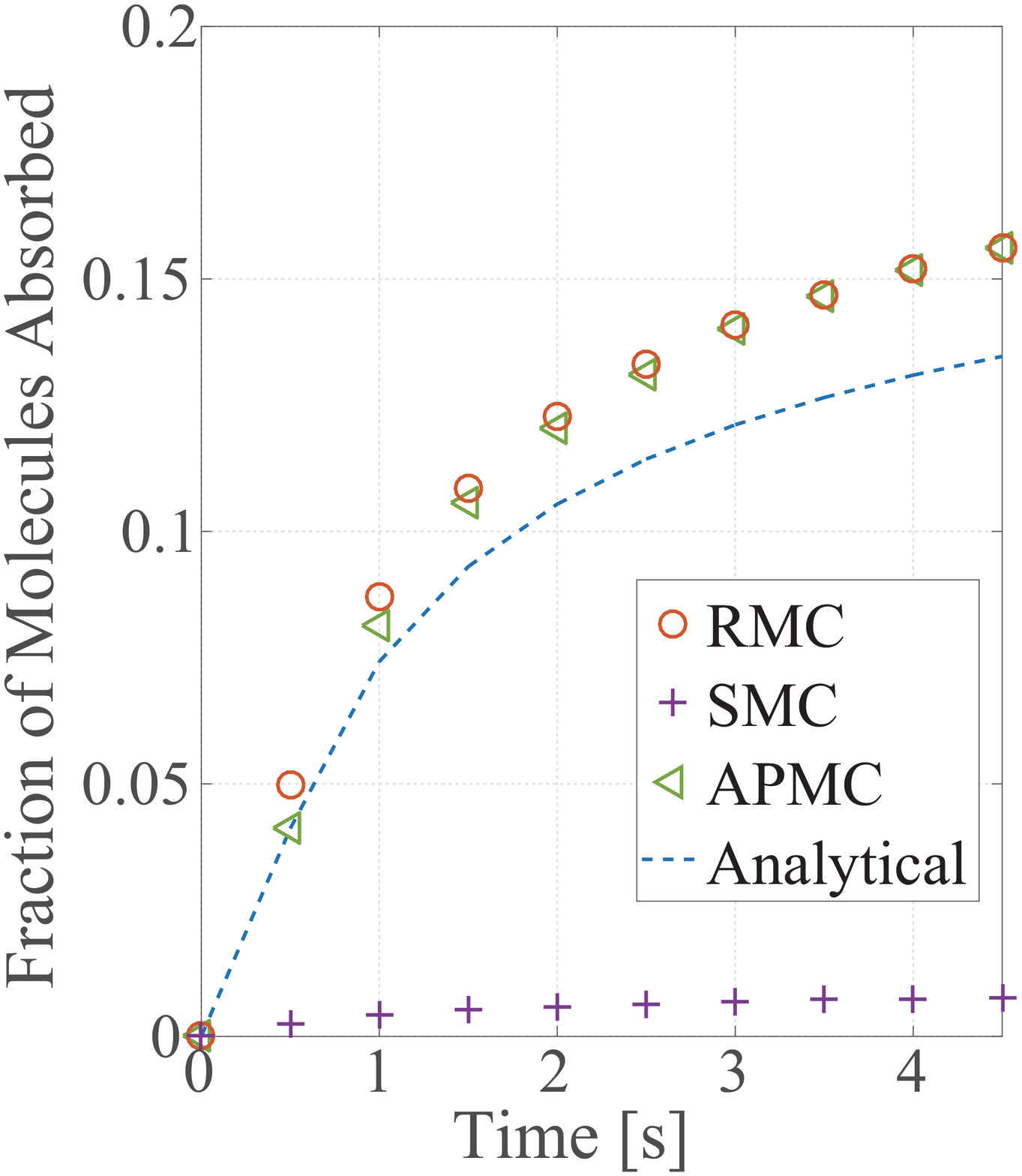}
    \caption{$\Delta t = 0.5\,\s$}\end{subfigure}
    \begin{subfigure}[b]{0.24\textwidth}\label{planar_0121}
    \includegraphics[width=\textwidth,height=2in]{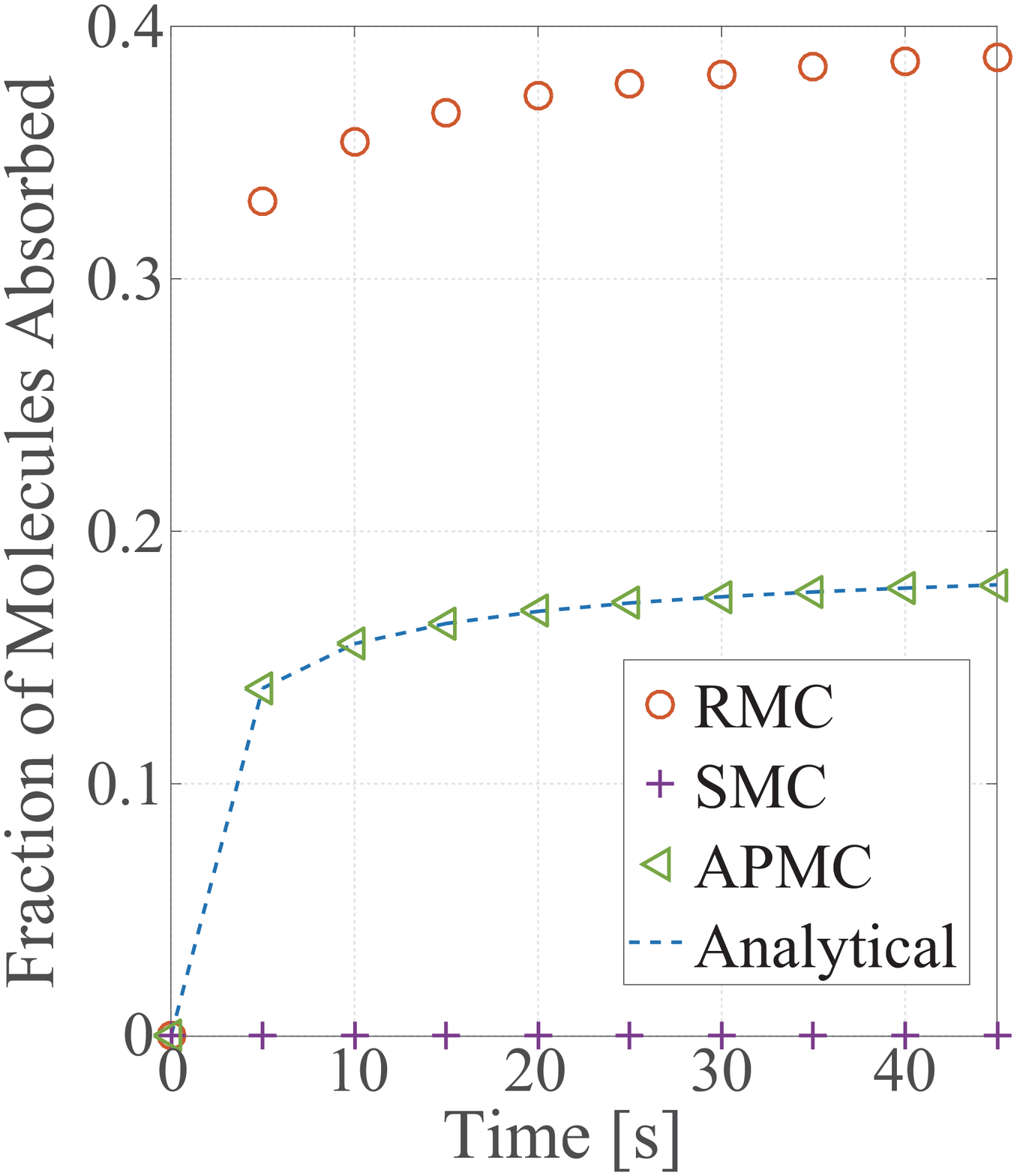}
    \caption{$\Delta t = 5\,\s$}\end{subfigure}
    \caption{Comparison of the time-varying results produced by the SMC algorithm, the RMC algorithm, and the APMC algorithm for $M = 100$ and $\radR = 10\,\mu \m$.\vspace{-1em}}\label{timevarying2}
\end{figure}

Fig.~\ref{timevarying2} plots the simulated fraction of molecules absorbed, together with the analytical result from \eqref{yilmazequation}, versus time $t$ for $M = 100$, $\radR = 10\,\mu\m$ and $\Delta t = 0.5\,\s$ or $5\,\s$. We observe from this figure that when $\Delta t$ increases from $0.5\,\s$ to $5\,\s$, our APMC algorithm achieves a higher accuracy. Specifically, Fig.~\ref{timevarying2}(a) shows that the gap between the fraction of molecules absorbed produced by the APMC algorithm and that produced by the RMC algorithm is very small. Also, Fig.~\ref{timevarying2}(a) shows that both the APMC and the RMC algorithms overestimate the fraction of molecules absorbed. Unlike Fig.~\ref{timevarying2}(a), Fig.~\ref{timevarying2}(b) shows that the APMC algorithm matches the fraction of molecules absorbed expected by the analytical result. Fig.~\ref{timevarying2}(b) shows that the fraction of molecules absorbed produced by the RMC algorithm is approximately twice of that produced by the APMC algorithm when $t\geq10\,\s$. This demonstrates the accuracy of our APMC algorithm when $\sqrt{D \Delta t}/\radR$ is large.

\begin{figure}[!t]
\centering
    \begin{subfigure}[b]{0.24\textwidth}
    \includegraphics[width=\textwidth,height=2in]{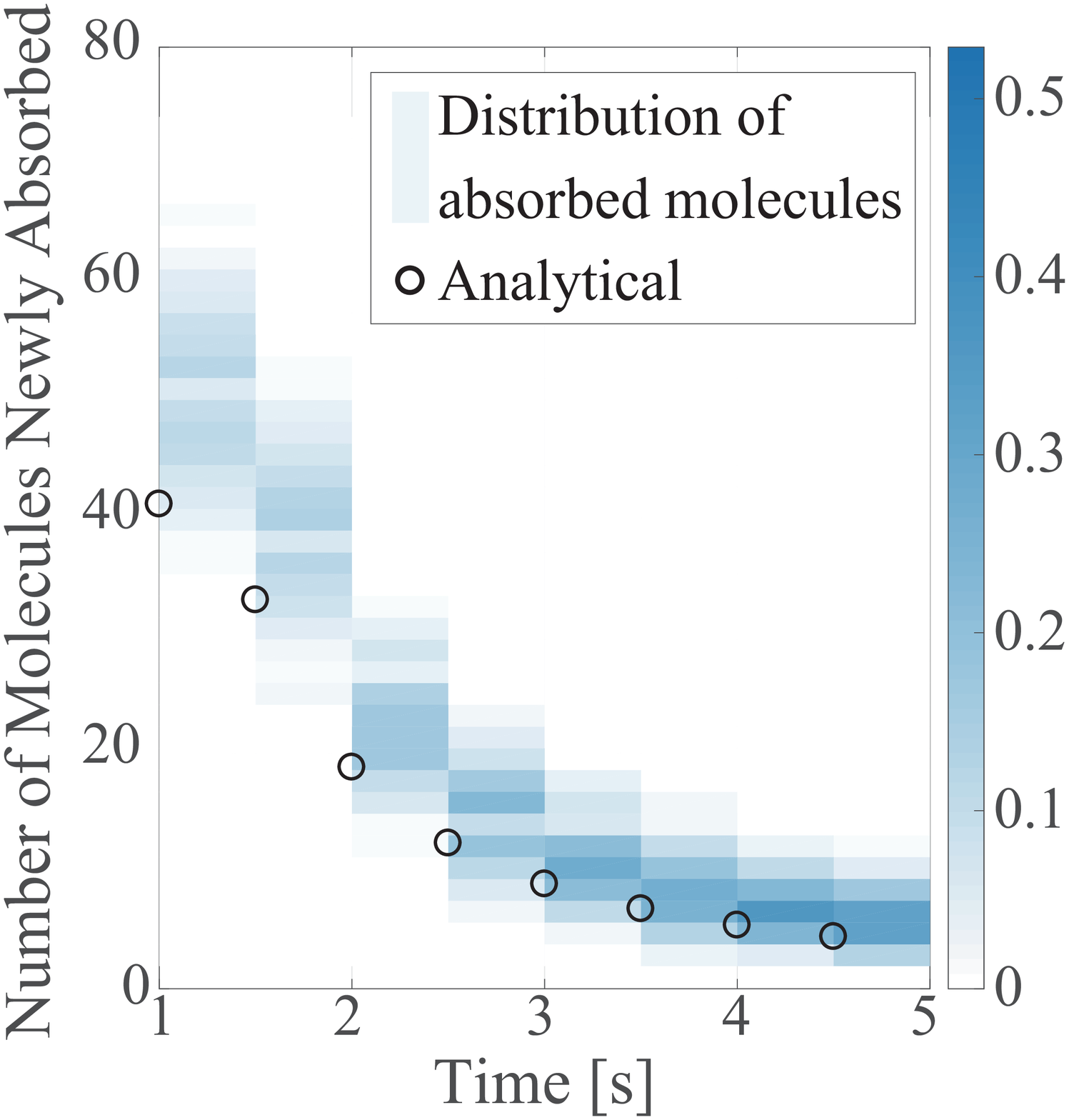}
    \caption{RMC Algorithm}\end{subfigure}
    \begin{subfigure}[b]{0.24\textwidth}
    \includegraphics[width=\textwidth,height=2in]{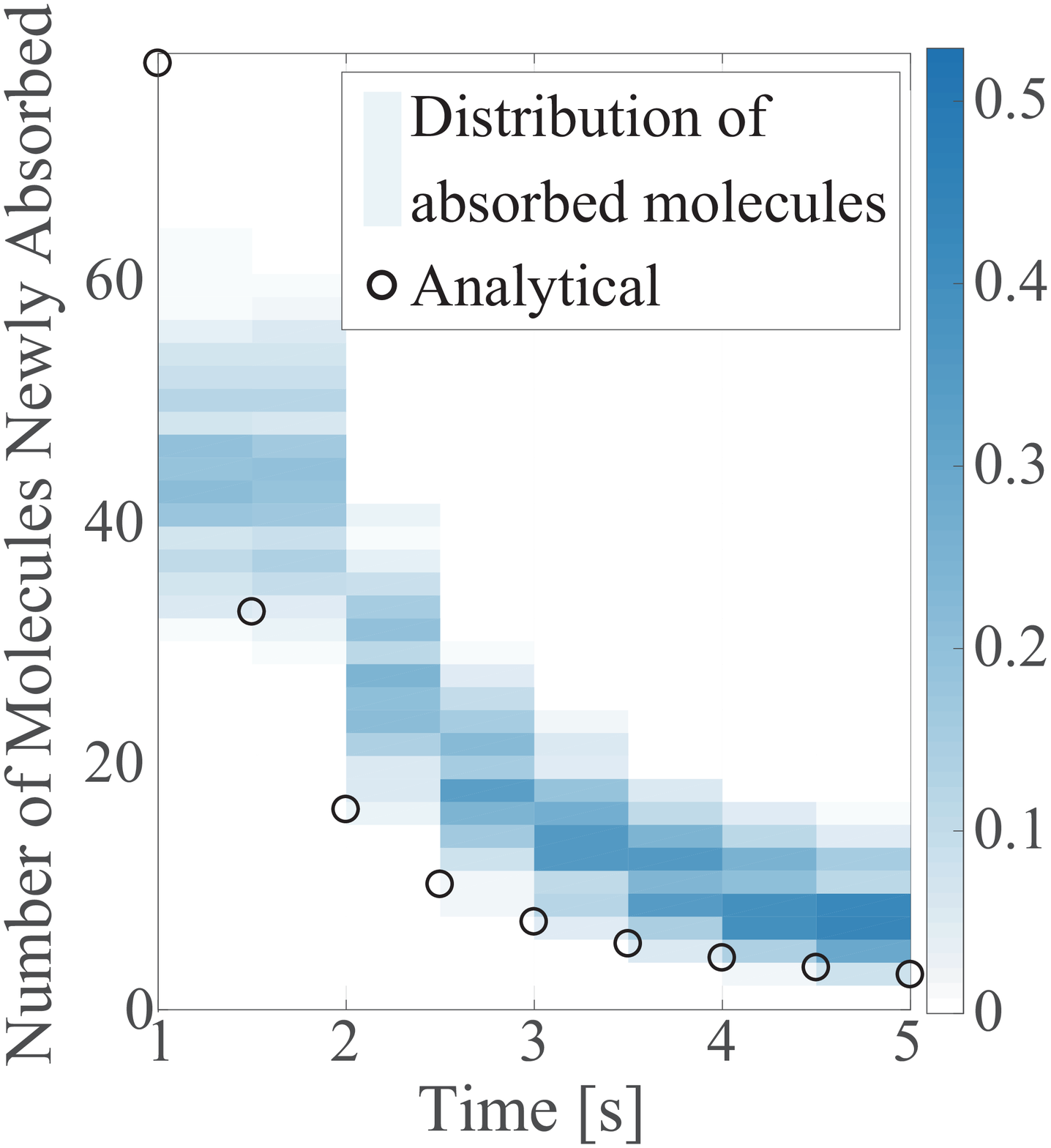}
    \caption{APMC Algorithm}\end{subfigure}
    \caption{Distribution of the number of molecules newly absorbed during each time step for the RMC and APMC algorithms with $\Delta t=0.5\,\s$, $M=10$, and $\radR=10\,\mu\m$. Simulations are repeated $10^{3}$ times and $N=10^{3}$ molecules are released each time. The spectrum bars represent observation probabilities.\vspace{-1em}}\label{Distribution}
\end{figure}

Fig.~\ref{Distribution} depicts the distribution of molecules newly absorbed during each time step (e.g., from $t=1\,\s$ to $t=1.5\,\s$) for both the RMC and APMC algorithms for $M = 10$, $\radR=10\mu m$, and $\Delta t=0.5\,\s$. The analytical result in this figure is obtained based on \eqref{yilmazequation}. We observe from this figure that the RMC algorithm significantly overestimates the number of molecules newly absorbed when $t=1\,\s$, while our APMC algorithm gives an improved estimation accuracy when $t=1\,\s$ by absorbing molecules according to \eqref{analysisequationtimestep}. When $t$ increases, the overestimation of the RMC algorithm is slightly less severe than that of the APMC algorithm.
We further observe that the distribution of molecules newly absorbed in Fig.~\ref{Distribution}(b) is very similar to that in Fig.~\ref{Distribution}(a), which demonstrates that our APMC algorithm does not noticeably affect this distribution. Importantly, the average of the simulated distribution using our APMC algorithm is not far from the analytical result.




\section{Conclusion}\label{sec:conclusion}

We proposed a new $\apriori$ Monte Carlo (APMC) algorithm to simulate the probability that a molecule is absorbed by a spherical receiver in MC systems. Based on numerical results, we confirmed that our APMC algorithm produces a more accurate result for the fraction of molecules absorbed than the existing RMC and SMC algorithms when $\sqrt{D\Delta t}/\radR$ is large. This demonstrates that our algorithm is suitable to enable an efficient simulation with a relatively large diffusion time step length.



\begin{thebibliography}{10}
\providecommand{\url}[1]{#1}
\csname url@samestyle\endcsname
\providecommand{\newblock}{\relax}
\providecommand{\bibinfo}[2]{#2}
\providecommand{\BIBentrySTDinterwordspacing}{\spaceskip=0pt\relax}
\providecommand{\BIBentryALTinterwordstretchfactor}{4}
\providecommand{\BIBentryALTinterwordspacing}{\spaceskip=\fontdimen2\font plus
\BIBentryALTinterwordstretchfactor\fontdimen3\font minus
  \fontdimen4\font\relax}
\providecommand{\BIBforeignlanguage}[2]{{%
\expandafter\ifx\csname l@#1\endcsname\relax
\typeout{** WARNING: IEEEtran.bst: No hyphenation pattern has been}%
\typeout{** loaded for the language `#1'. Using the pattern for}%
\typeout{** the default language instead.}%
\else
\language=\csname l@#1\endcsname
\fi
#2}}
\providecommand{\BIBdecl}{\relax}
\BIBdecl

\bibitem{nakanobook}
T.~Nakano, A.~W. Eckford, and T.~Haraguchi, \emph{Molecular
  Communication}.\hskip 1em plus 0.5em minus 0.4em\relax Cambridge University
  Press, 2013.

\bibitem{berg1993random}
H.~C. Berg, \emph{Random Walks in Biology}.\hskip 1em plus 0.5em minus
  0.4em\relax Princeton University Press, 1993.

\bibitem{Yilmaz_CL_2014}
H.~B. Yilmaz, A.~C. Heren, T.~Tugcu, and C.~B. Chae, ``Three-dimensional
  channel characteristics for molecular communications with an absorbing
  receiver,'' \emph{IEEE Commun. Lett.}, vol.~18, no.~6, pp. 929--932, June
  2014.

\bibitem{Deng_TMBMC_2015}
Y.~Deng, A.~Noel, M.~Elkashlan, A.~Nallanathan, and K.~C. Cheung, ``Modeling
  and simulation of molecular communication systems with a reversible
  adsorption receiver,'' \emph{IEEE Trans. Mol. Biol. Multi-Scale Commun.},
  vol.~1, no.~4, pp. 347--362, Dec. 2015.

\bibitem{ahmadzadeh2016comprehensive}
A.~Ahmadzadeh, H.~Arjmandi, A.~Burkovski, and R.~Schober, ``Comprehensive
  reactive receiver modeling for diffusive molecular communication systems:
  Reversible binding, molecule degradation, and finite number of receptors,''
  \emph{IEEE Trans. Nanobiosci.}, vol.~15, no.~7, pp. 713--727, Dec. 2016.

\bibitem{yilmaz2014simulation}
H.~B. Yilmaz and C.-B. Chae, ``Simulation study of molecular communication
  systems with an absorbing receiver: Modulation and {ISI} mitigation
  techniques,'' \emph{Simulation Model. Practice. Th.}, vol.~49, pp. 136--150,
  Dec. 2014.

\bibitem{llatser2014n3sim}
I.~Llatser, D.~Demiray, A.~Cabellos-Aparicio, D.~T. Altilar, and
  E.~Alarc{\'o}n, ``N3sim: Simulation framework for diffusion-based molecular
  communication nanonetworks,'' \emph{Simulation Model. Practice. Th.},
  vol.~42, pp. 210--222, Mar. 2014.

\bibitem{noel2017simulating}
A.~Noel, K.~C. Cheung, R.~Schober, D.~Makrakis, and A.~Hafid, ``Simulating with
  {AcCoRD}: Actor-based communication via reaction--diffusion,'' \emph{Nano
  Commun. Netw.}, vol.~11, pp. 44--75, Mar. 2017.

\bibitem{arifler2017}
D.~Arifler and D.~Arifler, ``Monte {C}arlo analysis of molecule absorption
  probabilities in diffusion-based nanoscale communication systems with
  multiple receivers,'' \emph{IEEE Trans. Nanobiosci.}, vol.~16, no.~3, pp.
  157--165, Apr. 2017.

\bibitem{andrews2004stochastic}
S.~S. Andrews and D.~Bray, ``Stochastic simulation of chemical reactions with
  spatial resolution and single molecule detail,'' \emph{Phys. Biol.}, vol.~1,
  no.~3, p. 137, Aug. 2004.

\bibitem{schulten2000lectures}
K.~Schulten and I.~Kosztin, ``Lectures in theoretical biophysics,''
  \emph{University of Illinois}, vol. 117, 2000.

\end{thebibliography}
\end{document}